\documentstyle[12pt]{article}
\def\dm{\displaystyle}\def\dm{\displaystyle}
\newcommand{\lie}{{\rm Lie\,}}
\newcommand{\const}{{\rm const\,}}
\def\dm{\displaystyle}
\renewcommand{\d}{{\rm d\,}}
\renewcommand{\det}{{\rm det\:}}

\begin{document}

\title{Quantum mechanics on Riemannian Manifold in Schwinger's
Quantization Approach III}

\author{Chepilko Nicolai Mikhailovich\\
\small\it
Physics Institute of the Ukrainian Academy of Sciences, 
Kyiv-03 028, Ukraine \\ \small\it e-mail: chepilko@zeos.net
\and
Romanenko Alexander Victorovich\\
\small\it
Kyiv Taras Shevchenko University,
Department of Physics, Kyiv-03 022, Ukraine\\
\small\it e-mail: ar@ups.kiev.ua}

\maketitle

\begin{abstract}

Using extended Schwinger's quantization approach quantum mechanics
on a Riemannian manifold $M$ with a given action of an intransitive group
of isometries is developed. It was shown that quantum mechanics
can be determined unequivocally only on submanifolds of $M$ where
$G$ acts simply transitively (orbits of $G$-action). The remaining part of
degrees of freedom can be described unequivocally after introducing
some additional assumptions. Being logically unmotivated, these assumptions
are similar to canonical quantization postulates. Besides this ambiguity
that has a geometrical nature there is undetermined gauge field of
$\hbar$ (or higher) order, vanishing in the classical limit $\hbar\to 0$.
\end{abstract}

\section{Introduction}

The purpose of the present paper is to continue a series of our
works devoted to a generalization of Schwinger's quantization
approach for the case of a Riemannian manifold with a group
structure (determined by its metric).

In this paper we turn to the most problematic case (in physical meaning)
of the formulation of quantum mechanics on a Riemannian manifold $M$
with the intransitive group of isometries $G$. Partially our consideration
is based on our previous results \cite{ourwork1}, \cite{ourwork2}.
But, in contradiction to them, the main feature of the present paper
is the fact that the dimension of the manifold $M$ is bigger then
the dimension of the group of isometries $G$. This leads to the decomposition
of degrees of freedom describing a point particle on $M$ into two sets,
related to the group $G$ and to the quotient space $M/G$. The last
manifold has not a global group structure in a general case. Due to
this reason it happens that the quantum mechanics is completely defined
only on submanifolds of $M$, which are isomorphic to $G$ (the orbits of
action of $G$ on $M$). The dynamical equations for the other degrees of
freedom, connected with $M/G$, can be unequivocally determined after
introducing some new assumptions expanding Schwinger's scheme from outside,
that are similar to canonical quantization postulates and can not be
motivated logically. Nevertheless, having introduces them, we find that in
quantum mechanics there is an abelian gauge field of $\hbar$ (or higher)
order, that remains undetermined and vanishes in the classical limit
$\hbar\to 0$.

\section{Structure of Manifold with Non-transitive Group of Isometries}

Let us consider $p$-dimensional Riemannian manifold $M$ equipped with the
metric $\{\eta_{MN}: M,N=\overline{1, p}\}$ in which the action of
$n$-dimensional intransitive group of isometries $G$ is given ($n<p$).
This means, according to \cite{eisen}, that the Killing equations
\begin{equation}
v^{P}\partial_{P}\eta_{MN}+\eta_{MP}\partial_{N} v^{P}+
\eta_{PN}\partial_{M} v^{P}=0
\label{1}
\end{equation}
have $n$ independent solutions
$\{v^{N}_{A}:\:N=\overline{1, p};\: A=\overline{1, n}\}$.

The set of vector fields $\{v^{M}_{A}\partial_{M}:
\:A=\overline{1, n}\}$ describes the representation of the Lie
algebra $\lie (G)$ of the Lie Group $G$ acting on $M$. Due to this
fact each the vector from this set obeys the equation
\begin{equation}
v^{N}_{A}\partial_{N} v^{M}_{B}-v^{N}_{B}\partial_{N} v^{M}_{A}=
C^{C}{}_{AB} v^{M}_{C}\,,
\label{2}
\end{equation}
where $C^{C}{}_{AB}$ are the structure constants of $G$.
This equality provides the following equation to have $m=p-n$
independent solutions
\begin{equation}
v^{N}_{A}\partial_{N}\varphi (x)=0\,.
\label{3}
\end{equation}

Let us consider a new coordinate system
\begin{equation}
\overline{x}^{N}:=(\varphi^{\alpha}(x), x^{\mu})
\label{4}
\end{equation}
$$
N=\overline{1, p}\,,\quad
\alpha=\overline{1, m}\,,\quad
\mu=\overline{m+1, p}
$$
where $\{\varphi^{\alpha}(x):\: \alpha=\overline{1, m}\}$
denote independent solutions of (\ref{3}). In these coordinates
the Killing vectors take the form:
\begin{equation}
\overline{v}^{N}_{A}=(v^{M}_{A}\partial_{N}\varphi^{\alpha},\: v^{\mu}_{A})
\equiv (0, v^{\mu}_{A})\,.
\label{5}
\end{equation}

Further we will assume that such a coordinate system denoted by
$\{x^{N}\}$ on $M$ is given and all the geometrical objects are
expressed in terms of it. The Killing vector can be rewritten as

\begin{equation}
v^{M}_{i}=\delta^{M}_{\mu}v_{i}^{\mu}\,,
\label{6}
\end{equation}
where the group index $i=\overline{m+1,\: p}$ is used instead of
$A=\overline{1,\: n}$ (the indices $i$, $j$, $k$, \dots play the same role
as $A$, $B$, $C$, \dots).

We introduce the inverse of the matrix
$\{v^{\mu}_{i}:\: \mu=\overline{m+1, p},\:i=\overline{m+1, p}\}$ as
\begin{equation}
e^{i}_{\mu}v^{\mu}_{j}=\delta^{i}_{j}\,,\quad
e^{i}_{\mu}v^{\nu}_{i}=\delta^{\nu}_{\mu}\,.
\label{7}
\end{equation}

Using (\ref{2}), (\ref{6}), (\ref{7}) one can prove that the matrix
$\{e^{i}_{\mu}\}$ obeys the Maurer-Cartan equation

\begin{equation}
\partial_{\mu}e_{\nu}^{i}-\partial_{\nu}e_{\mu}^{i}=
-C^{i}{}_{jk} e^{j}_{\mu}e^{k}_{\nu} \,.
\label{8}
\end{equation}

It is easy to obtain the following equations by differentiating

\begin{equation}
\begin{array}{rcl}
D_{M} v_{i}^{\mu}&=&\partial_{M}v_{i}^{\mu}+
\Lambda^{\mu}_{M\nu}v^{\nu}_{i}=0 \\[3mm]
D_{M}e^{i}_{\mu}&=&\partial_{M}e^{i}_{\mu}-
\Lambda^{\nu}_{M\mu}e^{i}_{\nu}=0
\end{array}
\label{10}
\end{equation}
where the object

\begin{equation}
\Lambda^{\mu}_{M\nu}=v^{\mu}_{i}\partial_{M}e^{i}_{\nu}
\label{11}
\end{equation}
corresponds to the right group connection on $M$ (see \cite{eisen}).

Using (\ref{7}), (\ref{8}) one can easily prove the following property
of $\Lambda^{\mu}_{M\nu}$:

\begin{equation}
\Lambda^{\sigma}_{\mu\nu}-\Lambda^{\sigma}_{\nu\mu}=
-C^{i}_{jk}v^{\sigma}_{i}e^{j}_{\mu}e^{k}_{\nu}\,.
\label{12}
\end{equation}

Using the relations obtained above we can write down the functional
features of the metric in the special coordinate system, described by (\ref{4}).
To do it we transform the Killing equation (\ref{1}) into new coordinate
description:

\begin{equation}
\begin{array}{rcl}
\partial_{\mu}\eta^{\alpha\beta}&=&0,\\[2mm]
\partial_{\mu}\eta_{\alpha\nu}&=&\eta_{\alpha\sigma}\Lambda^{\sigma}_{\nu\mu}
-\eta_{\nu\sigma}\Lambda^{\sigma}_{\alpha\mu,}\\[2mm]
\partial_{\mu}\eta_{\alpha\beta}&=&\eta_{\alpha\nu}\Lambda^{\nu}_{\beta\mu}
+\eta_{\beta\nu}\Lambda^{\nu}_{\alpha\mu}.
\end{array}
\label{13}
\end{equation}
The matrix $\{\eta_{\mu\nu}:\: \mu,\,\nu=\overline{m+1, p}\}$ have the sense
of the metric on the orbit of the action of $G$ on $M$, that is isomorphic to
$G$.

Making an assumption that $\{\eta_{\mu\nu}\}$  is non-degenerate matrix
we introduce the following objects:

\begin{equation}
A^{\mu}_{\alpha}=g^{\mu\nu}\eta_{\alpha\nu}\,,\quad
g^{\mu\nu}\eta_{\nu\sigma}=\delta^{\mu}_{\sigma}\,.
\label{14}
\end{equation}
Then $\eta_{\alpha\mu}=g_{\mu\nu}A^{\nu}_{\alpha}$ and

\begin{equation}
D_{\mu}A^{\nu}_{\alpha}=\partial_{\mu}A^{\nu}_{\alpha}
+\Lambda^{\nu}_{\mu\sigma}A^{\sigma}_{\alpha}=
\Lambda^{\nu}_{\alpha\mu}\,,
\label{15}
\end{equation}
as it follows from (\ref{13}).

Due to (\ref{14}), (\ref{15}) the second equation in (\ref{13})
is equivalent to

\begin{equation}
\partial_{\mu} (\eta_{\alpha\beta}-g_{\mu\nu}A^{\mu}_{\alpha}A^{\nu}_{\beta})=0\,.
\label{16}
\end{equation}
Hence, the matrix

\begin{equation}
g_{\alpha\beta}=\eta_{\alpha\beta}-g_{\mu\nu}A^{\mu}_{\alpha}A^{\nu}_{\beta}
\label{17}
\end{equation}
depends only on coordinates $\{x^{\alpha}:\: \alpha=\overline{1, m}\}$
and is independent on $\{x^{\mu}:\: \mu=\overline{m+1, p}\}$.
The metric tensor $\{\eta_{MN}\}$, describing the manifold $M$
can be rewritten in terms of objects $g_{\mu\nu}$, $A^{\mu}_{\alpha}$,
$g_{\alpha\beta}$ as follows

\begin{equation}
\{\eta_{MN}\}=
\left(
\begin{array}{cc}
g_{\alpha\beta}+g_{\rho\sigma}A^{\rho}_{\alpha}A^{\sigma}_{\beta}&
g_{\mu\rho}A^{\rho}_{\beta}\\ A^{\rho}_{\alpha}g_{\rho\nu}&
g_{\mu\nu}
\end{array}
\right)\,,
\label{18}
\end{equation}

\begin{equation}
\{\eta^{MN}\}=
\left(
\begin{array}{cc}
g^{\alpha\beta}&-A^{\mu}_{\gamma}g^{\gamma\beta}\\
-g^{\alpha\gamma}A^{\nu}_{\beta}&
g^{\mu\nu}+g^{\gamma\delta}A^{\mu}_{\gamma}A^{\nu}_{\delta}
\end{array}
\right)\,.
\label{19}
\end{equation}

This decomposition is similar to Kaluza-Klein one. In (\ref{19})
$\{g^{\alpha\beta}\}$ denotes the inverse of
$\{g_{\alpha\beta}\}$. If $\{\eta_{MN}\}$ and $\{g_{\mu\nu}\}$ are
non-degenerate matrices the matrix $\{g^{\alpha\beta}\}$ exists
due to the following property $$
\det\{\eta_{MN}\}=\det\{g_{\alpha\beta}\}\cdot\det\{g_{\mu\nu}\}
$$

The functional properties of the metric (\ref{18}) are determined by
the following equations

\begin{equation}
\left\{
\begin{array}{rcl}
\partial_{\mu}g_{\alpha\beta}&=&0\\[2mm]
\partial_{\rho}g_{\mu\nu}&=&g_{\mu\sigma}\Lambda^{\sigma}_{\rho\mu}+
g_{\nu\sigma}\Lambda^{\sigma}_{\rho\nu}\\[2mm]
\partial_{\mu} A^{\nu}_{\alpha}&=&
\Lambda^{\nu}_{\alpha\mu}-A^{\rho}_{\alpha}\Lambda^{\nu}_{\rho\mu}
\end{array}
\right.
\label{21}
\end{equation}

By its written form, the matrices (\ref{18}), (\ref{19}) correspond to
analogous ones in \cite{ourwork2}, but their meaning is different.

From the geometrical point of view the construction we are considering
corresponds to the principal bundle with a total space $M$, a base space
$M/G$ and a structure group $G$. The coordinates $\{x^{\alpha}\}$ from the
set $\{x^{M}\}=\{x^{\alpha}, x^{\mu}\}$ describes the local coordinate system
in $M/G$, the matrix $\{g_{\alpha\beta}\}$~--- the metric tensor of $M/G$.

The projection map of the principal bundle $(M/G)\,(M, G)$ has the form
$$
\begin{array}{ccc}
p:\: M &\longrightarrow& M/G\\ \{x^{\alpha}, x^{\mu}\}
&\longrightarrow& \{x^{\alpha}\}\,.
\end{array}
$$
The vector fields
\begin{equation}
\hat{D}_{i}=v_{i}^{\mu}\partial_{\mu} \,,\quad
\hat{D}_{\alpha}=\partial_{\alpha}-A^{\mu}_{\alpha}\partial_{\mu}
\label{23}
\end{equation}
form the bases of vertical and horizontal fields correspondingly.
One can observe it from

\begin{equation}
\d p (\hat{D}_{i}|_{\{x^{\alpha}, x^{\mu}\}})=0\,,\quad
\d p (\hat{D}_{\alpha}|_{\{x^{\alpha}, x^{\mu}\}})=\partial_{\alpha}
|_{\{x^{\alpha}\}}\,.
\label{24}
\end{equation}
The objects $\{A^{\mu}_{\alpha}\}$ defined above correspond to the
$\lie(G)$-valued connection 1-form on $M$:

\begin{equation}
\omega^{i}=e^{i}_{\mu} \left( \d x^{\mu}+A^{\mu}_{\alpha} \d
x^{\alpha} \right)\,. \label{25}
\end{equation}

Now we write down some useful relations for the Riemannian geometry on $M$,
which describe its features explicitly.

Let $\nabla_{\alpha}$ be a standard covariant derivative
constructed with the metric $\{g_{\alpha\beta}\}$. Define the
operator on $M/G$ that generalize the operator
$\hat\nabla_{\alpha}$ as

\begin{equation}
\hat{\nabla}_{\alpha}B_{\beta}=\hat{D}_{\alpha}B_{\beta}-\Gamma^{\gamma}{}_{\alpha\beta}
B_{\gamma} \label{26}
\end{equation}
for some vector field $B_{\alpha}$ on $M/G$ depending on coordinates
$\{x^{\mu}\}$. Then

\begin{equation}
[\hat\nabla_{\alpha}, \hat\nabla_{\beta}]B_{\gamma}=
[\hat{D}_{\alpha},
\hat{D}_{\beta}]B_{\gamma}-R^{\delta}{}_{\alpha\beta\gamma}B_{\delta}\,,
\label{27}
\end{equation}
where $R^{\delta}{}_{\alpha\beta\gamma}$ is the curvature tensor
for the metric $g_{\alpha\beta}$.

The explicit expression for the commutator $[\hat{D}_{\alpha}, \hat{D}_{\beta}]$
can be obtained from its action on a scalar function $f(x^{\alpha}, x^{\mu})$.
Performing the simple calculation we can observe that

\begin{equation}
[\hat{D}_{\alpha}, \hat{D}_{\beta}] f=
-(\partial_{\alpha}A^{\mu}_{\beta}-
\partial_{\beta}A^{\mu}_{\alpha})\partial_{\mu}f+
(A^{\mu}_{\alpha}\partial_{\mu}A^{\nu}_{\beta}-
A^{\mu}_{\beta}\partial_{\mu}A^{\nu}_{\alpha})\partial_{\nu}f\,.
\label{28}
\end{equation}
Using (\ref{21}) in (\ref{28}):

\begin{equation}
A^{\mu}_{\alpha}\partial_{\mu}A^{\nu}_{\beta}-
A^{\mu}_{\beta}\partial_{\mu}A^{\nu}_{\alpha}=
-C^{i}{}_{ik}A^{j}_{\alpha}A^{k}_{\beta}v^{\nu}_{i}+
(\partial_{\alpha}A^{\nu}_{\beta}-\partial_{\beta}A^{\nu}_{\alpha})+
(\partial_{\alpha}A^{i}_{\beta}-\partial_{\beta}A^{i}_{\alpha})v^{\nu}_{i}\,,
\label{29}
\end{equation}
where

\begin{equation}
A^{i}_{\alpha}=e^{i}_{\mu}A^{\mu}_{\alpha}\,,\quad
\partial_{\mu}A^{i}_{\alpha}=-C^{i}{}_{jk}e^{j}_{\mu}A^{k}_{\alpha}+
\partial_{\alpha}e^{i}_{\mu}\,.
\label{30}
\end{equation}

Due to (\ref{29}), (\ref{30}) the expression (\ref{28}) can be
rewritten as

\begin{equation}
[\hat{D}_{\alpha}, \hat{D}_{\beta}] f=
-F^{i}_{\alpha\beta}\hat{D}_{i} f\,,\quad
\hat{D}_{i}=v^{\mu}_{i}\partial_{\mu}\,.
\label{31}
\end{equation}

\begin{equation}
F^{i}_{\alpha\beta}=\partial_{\alpha}A^{i}_{\beta}-\partial_{\beta}A^{i}_{\alpha}+
C^{i}{}_{jk}A^{j}_{\alpha}A^{k}_{\beta}\,.
\label{32}
\end{equation}
Observing (\ref{31})-(\ref{32}) we can draw a conclusion that the
objects $\{A^{i}_{\alpha}\}$ can be interpreted as gauge fields
defined on $M/G$ with the strength tensor $F^{i}_{\alpha\beta}$.

The other relations between the basic fields have the form

$$
[\hat{D}_{i}, \hat{D}_{j}]=C^{k}{}_{ij}\hat{D}_{k}\,,\quad
[\hat{D}_{\alpha}, \hat{D}_{i}]=0\,.
$$

At the end of this section we consider classes of coordinate
transformations $x^{\mu}\to\overline{x}^{\mu}$ preserving the form of the
metric (\ref{18})-(\ref{19}) and its functional structure.  Evidently, one
such a class consists of coordinate transformations on the orbit of the
action of $G$:

\begin{equation}
\left\{
\begin{array}{rcl}
\overline{x}^{\alpha}&=&x^{\alpha}\\
\overline{x}^{\mu}&=&\overline{x}^{\mu}(x^{\nu})
\end{array}
\right.
\label{35}
\end{equation}

Under such a transformation $g_{\mu\nu}$, $A^{\mu}_{\alpha}$ behave as
a tensor and a covariant vector correspondingly (the geometric properties are
described by $\mu$, $\nu$, \dots). In the same time $g_{\alpha\beta}$
transforms as a scalar.

The coordinate change on $M/G$
\begin{equation}
\left\{
\begin{array}{rcl}
\overline{x}^{\alpha}&=&\overline{x}^{\alpha}(x^{\beta})\\
\overline{x}^{\mu}&=&x^{\mu}
\end{array}
\right.
\label{36}
\end{equation}
is of the same class.

The objects $g_{\alpha\beta}$, $A^{\mu}_{\alpha}$ transform as
a tensor and a contravariant vector correspondingly (the geometric properties are
described by $\alpha$, $\beta$, \dots), while  $g_{\alpha\beta}$
transforms as a scalar.

The third class of transformations is described by

\begin{equation}
\left\{
\begin{array}{rcl}
\overline{x^{\alpha}}&=&x^{\alpha}\\
\overline{x}^{\mu}&=&x^{\mu}+\varphi^{\mu}(x^{\alpha})
\end{array}
\right.
\label{37}
\end{equation}

The main geometrical objects transform under (\ref{37}) as

$$
\overline{g}_{\mu\nu} (\overline{x})=g_{\mu\nu}(x)\,,\quad
\overline{g}_{\alpha\beta} (\overline{x})=
g_{\alpha\beta}(x)\,,\quad
\overline{A}^{\mu}_{\alpha}(\overline{x})=
A^{\mu}_{\alpha}-\partial_{\alpha}\varphi^{\mu}(x^{\alpha})\,.
$$
An arbitrary vector $F_{M}$ on $M$ can be decomposed into two  objects
that are invariant under the transformation (\ref{37}):

$$
{\cal F}_{\alpha}=F_{\alpha}-A^{\mu}_{\alpha}F_{\mu}\,,\quad
{\cal F}_{\mu}=F_{\mu}\,.
$$
This decomposition means the extraction of the horizontal part of $F_{M}$.

\section{Lagrangian and Variational Principle}

Constructing the quantum mechanics on a Riemannian manifold $M$ with the
non-transitive group of isometries $G$ in terms of variational principle,
as in our previous papers, we assume that the coordinate operators $x^{M}$
form the total set of commuting observables.

The quantum Lagrangian can be written as
\begin{equation}
L=\frac{1}{2}\dot{x}^{M}\eta_{MN}(x)\dot{x}^{N}-U_{q}(x)\,,
\label{40}
\end{equation}
where $U_{q}(x)$ is some function that provides the scalar
transformation law of $L$ under a general non-degenerate
coordinate transformation $x\to \overline{x}=\overline{x}(x)$ (see
\cite{ourwork1}, \cite{ourwork2}). As it has been pointed out in
\cite{ourwork1}, its explicit expression can be determined in the
case when the commutator $[x^{M}, \dot{x}^{N}]$ is the function of
only $\{x^{M}\}$. This is a standard assumption in the formulation
of the quantum mechanics on a Riemannian manifold (see
\cite{sugano} and motivation in \cite{ourwork1}). The function
$U_{q}$ appears as the value of $\hbar^{2}$-order.

In the special coordinate system, introduced in (\ref{4}), the
metric tensor $\{\eta_{MN}\}$ receives the form that is similar to
one appearing in Kaluza-Klein theories. The Lagrangian can be
written as
\begin{equation}
L=\frac{1}{2}\left( \dot{x}^{\mu}+\dot{x}^{\alpha}A^{\mu}_{\alpha} \right)
g_{\mu\nu}
\left( \dot{x}^{\nu}+A^{\nu}_{\beta}\dot{x}^{\beta} \right)+
\frac{1}{2}\dot{x}^{\alpha}g_{\alpha\beta}\dot{x}^{\beta}-U_{q}(x)\,.
\label{41}
\end{equation}
The Euler-Lagrange dynamical equations, according to \cite{ourwork1},
can be derived from the action principle in the form of the variational
equation $\delta L=0$, that corresponds to the infinitesimal coordinate
variation $x^{M}\to x^{M}+\delta x^{M}(x)$.


As it has been developed in \cite{ourwork1}, the equality $\delta L=0$
holds if and only if the variations $\delta x^{M}$ are Killing vectors.
In the present case, in contradiction to one investigated in \cite{ourwork1}
and \cite{ourwork2}, the dimension of the manifold $M$ is less than the number
of linear independent Killing vectors
$\{v^{M}_{A}: M=\overline{1, p}\,,A=\overline{1, n}\}$ (i.e. $n<p$).

Extracting the total time derivative, as in \cite{ourwork1}, we can
rewrite the Lagrangian as
\begin{equation}
\delta L=\frac{d}{dt}\left( p_{M}\circ\delta x^{M} \right)
-\dot{p}_{M}\circ\delta x^{M}+\frac{1}{2}\dot{x}^{M}\left( \delta
x^{P}\partial_{P}\eta_{MN} \right)\dot{x}^{N}-\delta
x^{M}\partial_{M} U_{q}+ \frac{1}{2}\left[ \delta
x^{M},\frac{d}{dt}\left[\dot{x}^{N}, \eta_{MN}\right] \right]\,,
\label{42}
\end{equation}
where $p_{M}=\eta_{MN}\circ\dot{x}^{N}$ means the momentum operator on $M$.

In accordance with \cite{ourwork1}, the object
\begin{equation}
G=p_{M}\circ\delta x^{M}
\label{43}
\end{equation}
have the sense of the generator of permissible variations.

Further we decompose the variation $\delta x^{M}$ in terms of the
basis of linear independent Killing vectors (\ref{6}):
\begin{equation}
\delta x^{M}=v^{M}_{i}\varepsilon^{i}\,,\quad
\varepsilon^{i}=\const\,,
\label{44}
\end{equation}

This allow us to rewrite (\ref{43}) as

\begin{equation}
G=\left( p_{M}\circ v^{M}_{i} \right)\varepsilon_{i}=p_{i}\varepsilon^{i}\,,
\quad
p_{i}:=p_{M}\circ v^{M}_{i}\,.
\label{45}
\end{equation}
In the special coordinate system (\ref{4}) the objects $\{p_{i}\}$
have the form

\begin{equation}
p_{i}=p_{\mu}\circ v^{\mu}_{i}\,,\quad
p_{\mu}=\eta_{\mu M}\circ \dot{x}^{M}=g_{\mu\nu}\circ
\left( \dot{x}^{\nu}+A^{\nu}_{\alpha}\circ \dot{x}^{\alpha} \right)\,.
\label{46}
\end{equation}

\section{Algebra of Commutation Relations}

Following the procedure presented in \cite{ourwork1}, we derive the
commutation relations for quantum theory on $M$ performing investigation of
the properties of permissible variations connected with action of the group
$G$ of isometries on $M$. From the definition of permissible variations  in
terms of the representation of $G$ on $M$ (in the special coordinate system)
we have

\begin{equation}
\delta_{i}x^{\mu}=v^{\mu}_{i}=\frac{1}{i\hbar}\left[ x^{\mu}, p_{i} \right]\,.
\label{47}
\end{equation}

\begin{equation}
\delta_{i}x^{\alpha}=0=\frac{1}{i\hbar}\left[ x^{\alpha}, p_{i} \right]\,.
\label{48}
\end{equation}
(here $\delta x^{\mu}=\varepsilon^{i}\delta_{i} x^{\mu}$). Using
the assumption about the commutativity of the coordinates
$\{x^{M}\}$ and taking into account the fact that the matrix
$\{v^{\mu}_{i}\}$ is non-degenerate (i.e. $\det v^{\mu}_{i}\ne
0$), we can conclude from (\ref{47}), (\ref{48}) that

\begin{equation}
[x^{\mu}, p_{\nu}]=i\hbar\delta^{\mu}_{\nu}\,,\quad
[x^{\alpha}, p_{\mu}]=0\,.
\label{49}:
\end{equation}
The variation of the arbitrary function $f$ which depends on the
coordinates $\{x^{M}\}$ can be defined as follows

\begin{equation}
\delta_{i}f(x)=v_{i}^{\mu}\partial_{\mu}f(x)=\frac{1}{i\hbar}[f, p_{\mu}]\,.
\label{50}
\end{equation}
Hence

\begin{equation}
[f(x), p_{\mu}]=i\hbar\partial_{\mu}f(x)\,.
\label{51}
\end{equation}

The variations of the velocity operators are
\begin{equation}
\delta_{i}\dot{x}^{\mu}=\frac{d v^{\mu}_{i}}{d t}=
\dot{x}^{M}\circ \partial_{M}v^{\mu}_{i}=\frac{1}{i\hbar}[\dot{x}^{\mu}, p_{i}]\,,
\label{52}
\end{equation}

\begin{equation}
\delta_{i}\dot{x}^{\alpha}=\frac{d v^{\alpha}_{i}}{d t}=0=
\frac{1}{i\hbar}[\dot{x}^{\alpha}, p_{i}]\,.
\label{53}
\end{equation}

Using (\ref{45}), (\ref{46})  we can express the velocity
operators $\{\dot{x}^{\mu}\}$ in (\ref{52}) in terms of the
momentum operators $\{p_{\mu}\}$. Finally, we have

\begin{equation}
[p_{\mu}, p_{i}]=-i\hbar p_{\nu}\circ \partial_{\mu} v^{\nu}_{i}\,.
\label{54}
\end{equation}
The commutator (\ref{54}) can be presented in another form, taking
into account (\ref{2})

\begin{equation}
[p_{i}, p_{j}]=-i\hbar C^{i}{}_{ik}p_{k}\,.
\label{55}
\end{equation}
Similarly, due to (\ref{2}) and (\ref{51}), we can write

\begin{equation}
[p_{\mu}, p_{\nu}]=0\,.
\label{56}
\end{equation}
The commutators (\ref{49}), (\ref{51}), (\ref{55}) and (\ref{56})
completely define quantum mechanics on the orbit of the action of $G$
on $M$, that has been analyzed in \cite{ourwork1}.

Further, returning to the manifold $M/G$, we define the momentum
operator

\begin{equation}
p_{\alpha}=\eta_{\alpha M}\circ \dot{x}^{M}
\label{57}
\end{equation}
and the auxiliary operator

\begin{equation}
\pi_{\alpha}=g_{\alpha\beta}\circ\dot{x}^{\beta}=
p_{\alpha}-A^{\mu}_{\alpha}\circ p_{\mu}\,.
\label{58}
\end{equation}
Due to $\partial_{\mu}g_{\alpha\beta}=0$, the objects $p_{\mu}$ and $\pi_{\mu}$
are connected by the following commutation relations

\begin{equation}
[p_{\alpha}, p_{i}]=-i\hbar\circ\partial_{\alpha}v^{\mu}_{i}\,,\quad
[\pi_{\alpha}, p_{i}]=0\,.
\label{59}
\end{equation}

These are all the commutators which can be obtained immediately from
the generator of permissible variations (\ref{45}). Note, that presented
above commutation relations are form-invariant under a general coordinate
transformation $x^{M}\to \overline{x}^{M}=\overline{x}^{M}(x)$.
The derivation of the remaining commutators requires the usage
of operator equalities and introducing some additional assumptions.

It follows from the basic assumptions that the commutator between coordinate
and momentum operators on $M$ is the function of only $\{x^{M}\}$, i.e.

\begin{equation}
[x^{M}, p_{N}]=i\hbar B^{M}_{N}(x)\,.
\label{61}
\end{equation}
Due to the commutation relations obtained above the parts of the matrix $B$
is already defined, namely $B^{M}_{\nu}=\delta^{M}_{\nu}$. Then we can write

\begin{equation}
B=
\left(
\begin{array}{cc}
B^{\alpha}_{\beta}&B^{\mu}_{\beta}\\
0&\delta^{\mu}_{\nu}
\end{array}
\right)
\label{63}
\end{equation}
where the unknown functions $B^{\alpha}_{\beta}$,
$B^{\mu}_{\beta}$ can be written in the following form motivated
by the correspondence principle

\begin{equation}
B^{\alpha}_{\beta}=\delta^{\alpha}_{\beta}+b^{\alpha}_{\beta}\,,\quad
B^{\mu}_{\alpha}=b^{\mu}_{\alpha}\,.
\label{63a}
\end{equation}
The new unknown objects $b^{\alpha}_{\beta}$ and $b^{\mu}_{\alpha}$
in (\ref{63}) are the functions of $x$ of $\hbar^{2}$ (or higher) order.
To derive their operator properties we can use the commutator
of the structure equation

\begin{equation}
v^{\mu}_{i}\partial_{\mu}v^{\nu}_{j}-v^{\mu}_{j}\partial_{\mu}v^{\nu}_{i}=
C^{k}{}_{ij}v^{\nu}_{k}
\label{64}
\end{equation}
with the momentum operator $p_{\alpha}$. Using the relation

\begin{equation}
[v^{\mu}_{i}, p_{j}]=i\hbar\hat{D}_{j}v^{\mu}_{i}
\label{65}
\end{equation}
(the ``long derivative'' $\hat{D}_{i}$ was introduced in
(\ref{23}), (\ref{24})) we can rewrite (\ref{64}) as

\begin{equation}
[v^{\mu}_{i}, p_{j}]-[v^{\mu}_{j}, p_{i}]=
-C^{k}{}_{ij}v^{\mu}_{k}\,.
\label{66}
\end{equation}
Therefore, using the Jacobi identity, we arrive to the following operator
equality

\begin{equation}
[[v^{\mu}_{i}, p_{j}], p_{\alpha}]=
-\hbar^{2}\partial_{\nu}v^{\mu}_{i}\partial_{\alpha}v^{\nu}_{j}+
i\hbar\hat{D}_{j}[v^{\mu}_{i}, p_{\alpha}]\,.
\label{67}
\end{equation}

Further, taking the antisymmetrization of (\ref{67}) with respect to
the indices $i$, $j$ we can find the equation

\begin{equation}
\hat{D}_{i}\varphi^{\mu}{}_{j\alpha}-\hat{D}_{j}\varphi^{\mu}{}_{i\alpha}=
C^{k}{}_{ij}\varphi^{\mu}_{k\alpha}
\label{68}
\end{equation}
where we introduce the unknown function of coordinates defined as

\begin{equation}
\varphi^{\mu}{}_{i\alpha}=\frac{1}{i\hbar}[v^{\mu}_{i}, p_{\alpha}]-
\partial_{\alpha}v^{\mu}_{i}\,.
\label{69}
\end{equation}
Taking into account the structure of the operator $\hat{D}_{i}$ we
can conclude that the solution of (\ref{68}) has the form

\begin{equation}
\varphi^{\mu}{}_{i\alpha}=\hat{D}_{i}\varphi^{\mu}_{\alpha}\,,
\label{70}
\end{equation}
where $\varphi^{\mu}_{\alpha}$ is a new unknown function of $\{x^{M}\}$.

Making the substitution $p_{i}\equiv v^{\mu}_{i}\circ p_{\mu}$ into
(\ref{59}) we obtain

\begin{equation}
[p_{\alpha}, p_{\mu}]=i\hbar\partial_{\mu}\varphi^{\nu}_{\alpha}\circ p_{j} \,.
\label{71}
\end{equation}

Now we are going to show the relation between the objects $\varphi^{\mu}_{\alpha}$
and $B^{\nu}_{\alpha}$. To do it let us write down the Jacobi identities
including the coordinate operators $x^{M}$ and momentum operators
$p_{\alpha}$, $p_{\mu}$:

\begin{equation}
\frac{1}{i\hbar}[x^{\alpha}, [p_{\beta}, p_{\mu}]]\equiv 0=
\frac{1}{i\hbar}\left( [[x^{\alpha}, p_{\beta}], p_{\mu}]-
[[x^{\alpha}, p_{\mu}], p_{\beta}]\right)=
i\hbar\partial_{\mu}B^{\alpha}_{\beta}\,.
\label{72}
\end{equation}

\begin{equation}
\frac{1}{i\hbar}[x^{\mu}, [p_{\alpha}, p_{\nu}]]\equiv
i\hbar\partial_{\nu}\varphi^{\mu}_{\alpha}=
\frac{1}{i\hbar}\left( [[x^{\mu}, p_{\alpha}], p_{\nu}]-
[[x^{\mu}, p_{\nu}], p_{\alpha}]\right)=
i\hbar\partial_{\nu}B^{\mu}_{\alpha}\,.
\label{73}
\end{equation}
From (\ref{72}) we can draw a conclusion about the independence of
$B^{\alpha}_{\beta}$ (and, in consequence, $b^{\alpha}_{\beta}$) on $x^{\mu}$,
because

\begin{equation}
\partial_{\mu}B^{\alpha}_{\beta}=\partial_{\mu}b^{\alpha}_{\beta}=0\,.
\label{74}
\end{equation}
At the same time, it follows from (\ref{73}) that the differential
connection between $\varphi^{\mu}_{\alpha}$ and $B^{\mu}_{\alpha}$
(and, in consequence, $b^{\alpha}_{\beta}$) is

\begin{equation}
\partial_{\mu}\varphi^{\nu}_{\alpha}=\partial_{\mu}B^{\nu}_{\alpha}=
\partial_{\mu}B^{\nu}_{\alpha}\,.
\label{75}
\end{equation}

The object $\varphi^{\mu}_{\alpha}$ appears in all the formulae we deal with
under the derivative operator $\partial_{\mu}$. Hence, using (\ref{75})
we can identify $\varphi^{\mu}_{\alpha}$ with $b^{\mu}_{\alpha}$, i.e.
take

\begin{equation}
\varphi^{\mu}_{\alpha}(x)=b^{\mu}_{\alpha}(x)\,.
\label{76}
\end{equation}

Due to the structure equation the commutator for $v^{\mu}_{i}$ and $p_{\alpha}$
reads

\begin{equation}
\frac{1}{i\hbar}[v^{\mu}_{i}, p_{\alpha}]=\partial_{\alpha}v^{\mu}_{i}+
v^{\nu}_{i}\partial_{\nu}b^{\mu}_{\alpha}\,.
\label{77}
\end{equation}
On the other hand, in accordance with (\ref{61}), this commutator
is equal to

\begin{equation}
\frac{1}{i\hbar}[v^{\mu}_{i} p_{\alpha}]=
B^{M}_{\alpha}\partial_{M}v^{\mu}_{i}=\partial_{\alpha} v^{\mu}_{i}+
b^{M}_{\alpha}\partial_{M}v^{\mu}_{i}\,.
\label{78}
\end{equation}
By comparing the equations (\ref{77}) and (\ref{78}) we obtain the
equation for the objects $\{b^{M}_{\alpha}\}$:

\begin{equation}
v^{M}_{i}\partial_{M}b^{\mu}_{\alpha}-b^{M}_{\alpha}v^{\mu}_{i}=0\,,\quad
\partial_{\mu}b^{\alpha}_{\beta}=0\,.
\label{79}
\end{equation}

This equation has the transparent geometrical meaning. In accordance with
\cite{eisen}, the object $\{b^{M}_{\alpha}\}$ determines vector fields on
$M$, that generate the one-parametric group of coordinate transformations,
commuting with isometries (isometric transformations).
In a general case these transformations does not form the representation
of an $p-n$-dimensional group.

As the solutions of (\ref{79}), the objects are not determined unequivocally.
If $b^{M}_{\alpha}$ denotes the solution of (\ref{79}), its linear
combination

\begin{equation}
\overline{b}^{M}_{\alpha}(x)=\varepsilon^{\beta}_{\alpha}(x^{\gamma})
b^{M}_{\beta}(x)
\label{80}
\end{equation}
also obeys the equation (\ref{79}) for arbitrary functions
$\varepsilon^{M}_{\alpha}=\varepsilon^{M}_{\alpha}(x^{\gamma})$.

Therefore, the remaining commutation relations are

\begin{equation}
\begin{array}{rcl}
&[x^{\alpha}, p_{\beta}]=& i\hbar\left(
\delta^{\alpha}_{\beta}+b^{\alpha}_{\beta} \right)\,,\\[2mm]
&[x^{\mu}, p_{\beta}]=& i\hbar b^{\mu}_{\beta}\,,\\[2mm]
&[p_{\mu},p_{\alpha}]=& -i\hbar\partial_{\mu}b^{\nu}_{\alpha}\circ
p_{\nu}\,,
\end{array}
\label{81}
\end{equation}
where $b^{M}_{\alpha}$ obeys the equation (\ref{79}).

In order to construct the self-contained algebra of commutation
relations for quantum mechanics on $M$ we have to obtain the
explicit form of $[p_{\alpha}, p_{\beta}]$. In a general case we
can write

\begin{equation}
[p_{\alpha}, p_{\beta}]=i\hbar\left( F^{M}{}_{\alpha\beta}\circ p_{M}
+\varphi_{\alpha\beta} \right)\,,
\label{82}
\end{equation}
where $F^{M}{}_{\alpha\beta}$, $\varphi_{\alpha\beta}$ are unknown
tensors of coordinates on $M$. To determine these objects we use the
Jacobi identities for operators $x^{\alpha}$, $p_{\beta}$, $p_{\gamma}$
and $x^{\mu}$, $p_{\alpha}$, $p_{\beta}$. After simple calculation we find
that

\begin{equation}
F^{\gamma}{}_{\alpha\beta}B^{\delta}_{\gamma}=
-\left(
B^{\gamma}_{\alpha}\partial_{\gamma}B^{\delta}_{\beta}-
B^{\gamma}_{\beta}\partial_{\gamma}B^{\delta}_{\alpha}
\right)\,,
\label{83}
\end{equation}

\begin{equation}
F^{\mu}{}_{\alpha\beta}+B^{\mu}_{\gamma}F^{\gamma}_{\alpha\beta}=
-\left(
B^{M}_{\alpha}\partial_{M}B^{\mu}_{\beta}-
B^{M}_{\beta}\partial_{M}B^{\mu}_{\alpha}
\right)\,.
\label{84}
\end{equation}
These equations allow us to define $F^{M}_{\alpha\beta}$ as a function of
$B^{M}_{\alpha}$.

To investigate the functional properties of the object
$\varphi_{\alpha\beta}(x)$ it is convenient to introduce the following
operators

\begin{equation}
\overline{p}_{\alpha}=p_{\alpha}-b^{\mu}_{\alpha}\circ p_{\mu}\,.
\label{85}
\end{equation}

According (\ref{82}), the operators (\ref{85}) satisfy the
following commutation relations

\begin{equation}
[\overline{p}_{\alpha}, p_{\mu}]=0\,,\quad
[\overline{p}_{\alpha}, \overline{p}_{\beta}]=i\hbar
\left( F^{\gamma}{}_{\alpha\beta}\circ\overline{p}_{\gamma}+
\varphi_{\alpha\beta} \right)\,.
\label{86}
\end{equation}
From the relations

\begin{equation}
[p_{\mu}, [\overline{p}_{\alpha}, \overline{p}_{\beta}]]=0\,,\quad
\partial_{\mu}F^{\gamma}{}_{\alpha\beta}=0\,,
\label{87}
\end{equation}
obtained from (\ref{85}), (\ref{86}) we can conclude that

\begin{equation}
\partial_{\mu}\varphi_{\alpha\beta}=0\,,
\label{88}
\end{equation}
i. e. $\varphi_{\alpha\beta}$ are the functions only of the
coordinates on the quotient space $M/G$.

The further development of a theory can be performed only due to
additional assumptions about the structure of the matrix $B^{M}_{N}$,
that cannot be described in terms of the basic principles, having made a
base of our quantization scheme.

Let $B^{M}_{N}$ has the simplest form $B^{M}_{N}=\delta^{M}_{N}$.
In this case, as it follows from obtained above, the quantum mechanics
is described by

\begin{equation}
[x^{M}, p_{N}]=i\hbar\delta^{M}_{N}\,,\quad
[x^{M}, \dot{x}^{N}]=i\hbar\eta^{MN}\,.
\label{89}
\end{equation}

The relations (\ref{89}) can be divided into two parts. The first one
corresponds to quantum mechanics on the orbit,

\begin{equation}
[x^{\mu}, p_{\nu}]=i\hbar\delta^{\mu}_{\nu}\,,\quad
[x^{\mu}, x^{\nu}]=0\,,\quad
[p_{\mu}, p_{\nu}]=0\,,
\label{90}
\end{equation}
While the second one describes the quantum mechanics on the quotient space
$M/G$:

\begin{equation}
[x^{\alpha}, x^{\beta}]=0\,,\quad
[x^{\alpha}, \pi_{\beta}]=i\hbar\delta^{\alpha}_{\beta}\,,\quad
[\pi_{\alpha}, \pi_{\beta}]=i\hbar
\left( p_{\mu}\circ F^{\mu}{}_{\alpha\beta}+\varphi_{\alpha\beta} \right)\,.
\label{91}
\end{equation}
The remaining commutation relations, containing in (\ref{89}) have the
form

\begin{equation}
\begin{array}{lll}
&[x^{\alpha}, x^{\mu}]=0\,,& [\pi_{\alpha}, p_{\mu}]=
-i\hbar\partial_{\mu}A^{\nu}_{\alpha}\circ p_{\nu}\,,\\[2mm]
&[x^{\mu}, \pi_{\alpha}]=-i\hbar A^{\mu}_{\alpha}\,,& [x^{\alpha},
p_{\mu}]=0\,.
\end{array}
\label{92}
\end{equation}

It is essential, that the features of quantum mechanics on $M/G$
depend on the tensor $\varphi_{\alpha\beta}$, which cannot be
determined i our quantization scheme.This object can be viewed as
the strength tensor of some abelian gauge field. To show this, we
have to take into account that the commutator (\ref{82}) can be
rewritten in the simpler form due to the restriction
$B^{M}_{N}=\delta^{M}_{N}$, namely

\begin{equation}
[p_{\alpha}, p_{\beta}]=i\hbar\varphi_{\alpha\beta}(x^{\gamma})\,.
\label{93}
\end{equation}
Using (\ref{93}) we find from the Jacobi identities for the operators
$p_{\alpha}$, $p_{\beta}$, $p_{\gamma}$ that $\varphi_{\alpha\beta}$
obeys the relation

\begin{equation}
\partial_{\alpha}\varphi_{\beta\gamma}+
\partial_{\beta}\varphi_{\gamma\alpha}+
\partial_{\gamma}\varphi_{\alpha\beta}=0\,.
\label{94}
\end{equation}
Then, the evident solution of (\ref{94}) is

\begin{equation}
\varphi_{\alpha\beta}=\partial_{\alpha}A_{\beta}-\partial_{\beta}A_{\alpha}\,,
\label{95}
\end{equation}
where $A_{\alpha}$ is some unknown abelian gauge field.

\section{Quantum Lagrangian on $M$}

Following \cite{ourwork1} we introduce the following operator as the
Lagrangian of a point particle with the unit mass

\begin{equation}
L(x, \dot{x}):=\frac{1}{2}(\dot{x}, \dot{x})=\frac{1}{2}(p, p)\,.
\label{96}
\end{equation}
Here $(\cdot\,,\:\cdot)$ means the scalar product on $M$ that
is invariant under a general coordinate transformation $x\to\overline{x}(x)$.
Its constructing has been detally discussed in \cite{ourwork1}. In the present paper
we modify this definition for the case

\begin{equation}
[x^{M}, p_{N}]=i\hbar B^{M}_{N}(x)\,,
\label{97}
\end{equation}
where $B^{M}_{N}\ne\delta^{M}_{N}$. One can simply show that
$B^{M}_{N}$ is the tensor field on $M$ (and, consequently,
$b^{M}_{N}=B^{M}_{N}-\delta^{M}_{N}$ is the tensor field too).

Under a general coordinate transformation $x\to\overline{x}(x)$ on $M$
the momentum operator transforms as

\begin{equation}
p_{M}\to \overline{a}^{N}_{M}\circ p_{N}=
\overline{a}^{N}_{M}p_{N}+\frac{1}{2}[p_{N},
\overline{a}^{N}_{M}]= \overline{a}^{N}_{M}p_{N}-
\frac{i\hbar}{2}(\partial_{N}\overline{a}^{N}_{M}+b^{P}_{M}\partial_{P}\overline{a}^{N}_{M})\,,
\label{98}
\end{equation}

\begin{equation}
p_{M}\to \overline{a}^{N}_{M}\circ p_{N}=
p_{N}\overline{a}^{N}_{M}-\frac{1}{2}[p_{N},
\overline{a}^{N}_{M}]= p_{N}\overline{a}^{N}_{M}+
\frac{i\hbar}{2}(\partial_{N}\overline{a}^{N}_{M}+b^{P}_{M}\partial_{P}\overline{a}^{N}_{M})\,,
\label{99}
\end{equation}

Taking into account the transformation law of the Christoffel symbols
$\Gamma^{P}_{MN}$, constructed with the metric $\eta_{MN}$, we can
write down the derivatives of the transformation matrices as

\begin{equation}
\partial_{N}\overline{a}^{N}_{M}=
\overline{\Gamma}_{M}+\overline{a}^{N}_{M}\Gamma_{N}\,,\quad
\Gamma_{N}=\Gamma^{M}_{MN}\,,
\label{100}
\end{equation}

\begin{equation}
\partial_{N}\overline{a}^{P}_{M} =
\overline{\Gamma}^{S}_{LM}\overline{a}^{L}_{N}\overline{a}^{P}_{S}-
\overline{a}^{S}_{M}\Gamma^{P}_{NS}\,.
\label{101}
\end{equation}
Let us introduce the notation

\begin{equation}
W_{M}=\Gamma^{L}_{MN}b^{N}_{L}\,.
\label{102}
\end{equation}
This object transforms under a general coordinate transformation on $M$ as

\begin{equation}
\overline{W}_{M}=\overline{a}^{N}_{M}W_{N}+b^{N}_{L}\partial_{N}
\overline{a}^{L}_{M}\,.
\label{103}
\end{equation}
Hence, taking into account the transformation laws of $p_{M}$, $\Gamma_{M}$
and $W_{M}$ we can define the following ``left'' and ``right'' parts of the
momentum operator

\begin{equation}
\begin{array}{rcl}
\pi_{M}&=&p_{M}-\dm\frac{i\hbar}{2}(\Gamma_{M}+W_{M})\,,\\[3mm]
\pi_{M}^{\dag}&=&p_{M}+\dm\frac{i\hbar}{2}(\Gamma_{M}+W_{M})\,,
\end{array}
\label{104}
\end{equation}
that transform as

$$
\pi^{\dag}_{M}\to\overline{\pi}^{\dag}_{M}=\overline{a}^{N}_{M}\pi^{\dag}_{N}\,,\quad
\pi_{M}\to\overline{\pi}_{M}=\pi_{N}\overline{a}^{N}_{M}\,.
$$

Following \cite{ourwork1} we define the scalar norm of the momentum
operator $p_{M}$ on $M$ in terms of the operators $\pi^{\dag}_{M}$
and $\pi_{M}$ as

\begin{equation}
(p, p)=\pi_{M}\eta^{MN}\pi^{\dag}_{N}\,.
\label{105}
\end{equation}

Using the obtained above, the Lagrangian can be written in the form

\begin{equation}
L=\frac{1}{2}(\pi_{\alpha}-\pi_{\mu}A^{\mu}_{\alpha})
g^{\alpha\beta}
(\pi^{\dag}_{\beta}-A^{\mu}_{\beta}\pi^{\dag}_{\beta})+
\frac{1}{2}\pi_{\mu}g^{\mu\nu}\pi^{\dag}_{\nu}\,.
\label{106}
\end{equation}
In the simplest case $B^{M}_{N}=\delta^{M}_{N}$ we can write it as

\begin{equation}
L=\frac{1}{2}(p_{M}-\frac{i\hbar}{2}\Gamma_{M}) \eta^{MN}
(p_{M}+\frac{i\hbar}{2}\Gamma_{M})=
\frac{1}{2}p_{M}\eta^{MN}p_{N}+\frac{\hbar^{2}}{4} \left(
\partial_{M}\Gamma^{M}+\frac{1}{2}\Gamma_{M}\Gamma^{M} \right)\,,
\label{107}
\end{equation}
where

\begin{equation}
\Gamma_{M}=\Gamma^{N}_{MN}\,,\quad
\Gamma^{M}=\eta^{MN}\Gamma_{N}\,.
\label{107a}
\end{equation}

The further analyzes will be performed in the restriction $B^{M}_{N}=\delta^{M}_{N}$.
Define the metric on the group manifold $G$:

\begin{equation}
\eta_{ij}=g_{\mu\nu}v^{\mu}_{i}v^{\nu}_{j}\,,\quad
\partial_{M}\eta_{ij} =0\,.
\label{108}
\end{equation}

The norm of the momentum operator on $G$

$$
{\cal P}|_{G}=\{v^{\mu}_{i}\circ p_{\mu}\,:\: i=\overline{m=1,\,p}\}
$$
is defined as

\begin{equation}
({\cal P}|_{G},\, {\cal P}|_{G})=p_{i}\eta^{ij}p_{j}\,.
\label{109}
\end{equation}
From the Killing equations in  the form $\nabla_{M}v^{M}=0$ it
follows that

\begin{equation}
\partial_{M}v^{M}_{A}\equiv\partial_{\mu}v^{\mu}_{A}=-\Gamma_{\mu}v^{\mu}_{A}\,.
\label{110}
\end{equation}
In the same time  we conclude from (\ref{11}) that

\begin{equation}
\Gamma_{\mu}=-e^{i}_{\mu}\partial_{\nu}v^{\nu}_{i}=
v^{\nu}_{i}\partial_{\nu}e^{i}_{\mu}=\Lambda^{\nu}{}_{\nu\mu}\,.
\label{111}
\end{equation}
Performing a direct calculation, one can obtain that

\begin{equation}
\Gamma_{\alpha}=\Gamma^{M}_{N\alpha}=
\frac{1}{2}g^{\beta\gamma}\partial_{\alpha}g_{\beta\gamma}+
\frac{1}{2}g^{\mu\nu}\partial_{\alpha}g_{\mu\nu}\,,
\label{112}
\end{equation}
Finally, due to (\ref{111}), (\ref{112}) the definition
(\ref{109}) can be written as

\begin{equation}
({\cal P}|_{G},\, {\cal P}|_{G})= \left(
p_{\mu}-\frac{i\hbar}{2}\Gamma_{\mu} \right)g^{\mu\nu} \left(
p_{\mu}+\frac{i\hbar}{2}\Gamma_{\mu} \right)\,,
\label{113}
\end{equation}
i. e. the norm of a group momentum coincides with the double Lagrangian
for a free particle on the orbit of $G$ on $M$.

Using (\ref{104}) we can transform the Lagrangian into another
form

\begin{equation}
L=\frac{1}{2}p_{j}g^{ij}p_{j}+ \frac{1}{2} \left(
p_{\alpha}-p_{\mu}A^{\mu}_{\alpha}-\frac{i\hbar}{2}\hat{\Gamma}_{\alpha}
\right) g^{\alpha\beta} \left(
p_{\alpha}-A^{\mu}_{\alpha}p_{\mu}+\frac{i\hbar}{2}\hat{\Gamma}_{\alpha}
\right)\,,
\label{113a}
\end{equation}
where

$$
\hat{\Gamma}_{\alpha}=\Gamma_{\alpha}-A^{\mu}_{\alpha}\Gamma_{\mu}
\,. $$ So we can rewrite (\ref{113}) in terms of
$\pi_{\alpha}=p_{\alpha}-A^{\mu}_{\alpha}\circ p_{\mu}$ as

\begin{equation}
L=\frac{1}{2}p_{i}\eta^{ij}p_{j}+ \frac{1}{2}\left(
\pi_{\alpha}-\frac{i\hbar}{2}\Omega_{\alpha} \right)
g^{\alpha\beta} \left(
\pi_{\alpha}+\frac{i\hbar}{2}\Omega_{\alpha} \right)\,,
\label{114}
\end{equation}
where

$$
\Omega_{\alpha}=\hat{\Gamma}_{\alpha}-\partial_{\mu}A^{\mu}_{\alpha} \,.
$$

In order to simplify (\ref{114}) we have to present the object
$\Omega_{\alpha}$ in the explicit form. Using the structure
equation we obtain

\begin{equation}
\partial_{\mu}A^{\mu}_{\alpha}=-A^{\nu}_{\alpha}\Lambda^{\mu}_{\nu\mu}+
\Lambda^{\mu}_{\alpha\mu}\,,
\label{115}
\end{equation}
and, consequently,

\begin{equation}
\Omega_{\alpha}=\Gamma_{\alpha}+A^{\mu}_{\alpha}\Lambda^{\nu}_{[\mu\nu]}-
\Lambda^{\mu}_{\alpha\mu}=\Gamma_{\alpha}+v^{\mu}_{i}\partial_{\alpha}e^{i}_{\mu}+
C^{i}{}_{ij}A^{j}_{\alpha}\,,
\label{116]}
\end{equation}
where

$$
A^{i}_{\alpha}=e^{i}_{\mu}A^{\mu}_{\alpha}\,.
$$

Further, taking into account (\ref{19}) and the equality

\begin{equation}
\det \{\eta_{AB}\}=\det \{g_{\alpha\beta}\}\cdot\det
\{v^{\mu}_{i}\}\,,
\label{117}
\end{equation}
we arrive to the final expression for $\Omega_{\alpha}$:

\begin{equation}
\Omega_{\alpha}=\gamma_{\alpha}+\frac{1}{2}\eta^{ij}\partial_{\alpha}\eta_{ij}+
C^{i}{}_{ij}A^{j}_{\alpha}\,,
\label{118}
\end{equation}
where $\gamma_{\alpha}=\gamma^{\beta}{}_{\alpha\beta}$,
$\gamma^{\alpha}{}_{\beta\gamma}$ denotes the Christoffel symbol
constructed with the metric $g_{\alpha\beta}$ of the quotient
space.

The formulae (\ref{114}) and (\ref{118}) completely determine
the final expression for the Lagrangian for freely moving particle
on $M$ in the restriction $B^{M}_{N}=\delta^{M}_{N}$. The first
term in (\ref{114}) corresponds to quantum theory on the orbits
of the action of $G$ on $M$, While the second one describes the
theory on the quotient space $M/G$. It is essential to point out
here  that the degrees of freedom $\{x^{\mu}\}$ determine the
quantum mechanics on $M/G$ by means of the last terms in
(\ref{114}).

\section{Equations of Motion for Dynamics on Riemannian Manifold}

In this section we restrict the function $B^{M}_{N}$ to be equal
to $\delta^{M}_{N}$. The way of determining the dynamical
equations of motion describing the particle moving on the manifold
with an intransitive group of isometries differs from one
introduced in \cite{ourwork1} only in details. From the condition
$\delta L=0$ of permissible variations we find that

\begin{equation}
v^{M}_{A}\circ\left(
\dot{p}_{M}+\frac{1}{2}p_{L}\partial_{M}\eta^{LN}p_{N}+\partial_{M}U
\right)=0\,, \label{119}
\end{equation}
here

\begin{equation}
U=\frac{\hbar^{2}}{4}\left( \partial_{M}\Gamma^{M}+
\frac{1}{2}\Gamma_{M}\Gamma^{M}\right)
\label{120}
\end{equation}
where we have substituted the Lagrangian (\ref{107}).

Using the condition $v^{\mu}_{i}\, e^{i}_{\nu}=0$, where
$\{e^{i}_{\mu}\}$ is the inverse of $\{v^{\mu}_{i}\}$ we obtain
the equations of motion of the free particle for degrees of
freedom connected with the orbits of the action of $G$ on $M$ in
the following form

\begin{equation}
\dot{p}_{\mu}=-\frac{1}{2}p_{M}\partial_{\mu}\eta^{MN}p_{N}-\partial_{\mu}U\,.
\label{121}
\end{equation}
This equation is equivalent to the conservation law
$\dot{p}_{i}=0$ for the group momentum $p^{i}$.

As to the other degrees of freedom associated with $M/G$, its dynamical
equations in Euler-Lagrange form must be equivalent to ones in Heisenberg
form

\begin{equation}
\dot{p}_{M}=\frac{1}{i\hbar}[p_{M}, H]\,.
\label{122}
\end{equation}

The  explicit expression of the Hamiltonian can be obtained by
comparison (\ref{121}) with (\ref{122}) taking $M:=\mu$. The
Hamiltonian operator derived in such a way has the following form

\begin{equation}
H=\frac{1}{2}p_{M}\eta^{MN}p_{N}+U+u_{q}(x^{\alpha})\,,
\label{123}
\end{equation}
where $u_{q}=u_{q}(x^{\alpha})$ is some function that does not
appear in (\ref{121}) (because $\partial_{\mu} u_{q}\equiv 0$).
The detail calculation allow us to put $u_{q}=0$ (see the remark
below).

Hence, we can transform (\ref{119}) into the following form

\begin{equation}
\dot{p}_{M}=-\frac{1}{2}p_{L}\partial_{M}\eta^{LN}p_{N}-
\partial_{M}U-\left( \eta^{LN}\circ\varphi_{MN} \right)\circ p_{L}\,,
\label{124}
\end{equation}
where

$$ \varphi_{MN}=\frac{1}{i\hbar}[p_{M}, p_{N}]\,. $$ The
restriction of $M$ to $\mu$ reduces (\ref{124}) to (\ref{121}),
as it must be.

To prove the fact that (\ref{124}) is correctly defined, we
consider the operator $H$ as the generator of time shifts $t\to
\overline{t}=t+\delta t(t)$, where $\delta t(t)$ are permissible
time variations. The result of the time variation of $L$ must be
in agreement with the equations of motion in Euler-Lagrange form
(\ref{124}).

The time shift causes the following variations of coordinate and velocity
operators

\begin{equation}
\begin{array}{rcl}
\delta x&=&\overline{x}(\overline{t})-x(t)=\dot{x}(t)\,,\\
\delta\dot{x}&=&\dm\frac{d}{dt}\delta x-\dot{x}\dm\frac{d\delta
t}{dt}\,.
\end{array}
\label{125}
\end{equation}
Using the commutation relations we obtain

\begin{equation}
[x^{M}, \delta x^{N}]=i\hbar\eta^{MN}\delta t\,, \quad
[\dot{x}^{M}, \delta x^{N}]=[\dot{x}^{M}, \dot{x}^{N}]\delta t\,,
\label{126}
\end{equation}
where

\begin{equation}
[\dot{x}^{M}, \dot{x}^{N}]=i\hbar\varphi^{MN}+i\hbar
f^{MN}_{L}\circ \dot{x}^{L}\,,
\label{127}
\end{equation}

\begin{equation}
\begin{array}{rcl}
\varphi^{MN}&=&\eta^{ML}\eta^{NS}\varphi_{LS}\,,\\
f^{MN}{}_{P}&=&\left( \eta^{MS}\eta^{NL}-\eta^{NS}\eta^{ML} \right)
\partial_{S}\eta_{LP}\,.
\end{array}
\label{128}
\end{equation}
The variation of the Lagrangian caused only by the time shift appears
in the variation of the action functional in the following combination

\begin{equation}
\delta W=\int\limits_{t_{2}}^{t_{1}} \delta_{t} L d\,t\,,\quad
\delta_{t}L=\delta L+L\frac{d\delta t}{dt}\,.
\label{129}
\end{equation}
The expression $\delta_{t}L$ reduces to

\begin{equation}
\delta_{t}L=\frac{d(L\delta t)}{dt}\,.
\label{130}
\end{equation}
only due to the commutation relations.

On the other hand, extracting the total time derivative we obtain

\begin{equation}
\delta L=\frac{d}{dt}\left( H\delta t \right)+
\frac{dL}{dt}-
\left( \dot{p}_{M}-\frac{1}{2}p_{N}\partial_{M}\eta^{NL}p_{L}-
\partial_{M}U -p_{N}\circ \varphi^{\mu}{}_{M} \right)\circ \delta x^{N}\,,
\label{131}
\end{equation}
here $H$ means the Hamiltonian. Hence, we can write

\begin{equation}
H=L\,,\quad
\frac{dH}{dt}=0\,,
\label{132}
\end{equation}
after comparing (\ref{130}) with (\ref{131}). Note that if we
define the Hamiltonian with an additional term $u_{q}(x)$, we
obtain after such a comparison that $u_{q}=0$ (otherwise the
equation (\ref{130}) and (\ref{131}) are not agreed). Therefore,
we can see that the equations (\ref{124}) are consistent with
other conclusions.

The equations of motion for degrees of freedom representing the orbits of the
action of $G$ on $M$ are introduced above (see (\ref{121})). To write down
the other dynamical equations describing the motion on $M/G$ it is sufficient
to assume $M\to \alpha\in\overline{1, m}$. In terms of $\pi_{\alpha}$
they are found in the form

\begin{equation}
\begin{array}{rll}
\dot{\pi}_{\alpha}&=&
-\dm\frac{1}{2}\pi_{\beta}\partial_{\alpha}g^{\beta\gamma}\pi_{\gamma}+
\dm\frac{1}{2}\left(
\pi_{\beta}g^{\beta\gamma}F^{\mu}{}_{\alpha\gamma}p_{\mu}+
p_{\mu}g^{\beta\gamma}F^{\mu}{}_{\alpha\gamma}\pi_{\beta}\right)\\[3mm]
&&-\dm\frac{1}{2} p_{\mu}\left(
g^{\mu\sigma}\partial_{\sigma}A^{\nu}_{\alpha}+
g^{\nu\sigma}\partial_{\sigma}A^{\mu}_{\alpha} \right) p_{\nu}+
p_{M}\circ\varphi^{M}_{\alpha}-\hat{D}_{\alpha}U\\[3mm]
&&+\dm\frac{\hbar^{2}}{4}\hat{D}_{\gamma}\left(\partial_{\alpha}g^{\beta\gamma}\partial_{\mu}A^{\mu}_{\beta}\right)
-\dm\frac{\hbar^{2}}{8}\partial_{\alpha}g^{\beta\gamma}\left(
\partial_{\mu}A^{\mu}_{\beta}\partial_{\nu}A^{\nu}_{\gamma}\right)\\[3mm]
&&+\dm\frac{\hbar^{2}}{4}\partial_{\mu}\left(
F^{\mu}{}_{\alpha\beta}g^{\beta\gamma}\partial_{\nu}A^{\nu}_{\gamma}\right)
+\dm\frac{\hbar^{2}}{4}\partial_{M}\left(
\eta^{MN}\partial_{M}\partial_{\mu}A^{\mu}_{\alpha} \right)\,.
\end{array}
\label{133}
\end{equation}

The equations (\ref{121}), (\ref{133}) completely describe the
motion of a particle on the manifold with an intransitive group of
motions. It appears that dynamics is decomposed into two parts.
The first one corresponds to the motion on the orbits of the
action of $G$  on $M$, the equations of motion have the form of
conservation laws for the group momenta
$\{p_{i}:\:i=\overline{m+1, p}\}$.  The second one is related to
the motion on $M/G$, where the dynamics of a particle is governed
by Lorentz-type forces. The first one is caused by non-abelian
gauge field $A^{\mu}_{\alpha}$ with the strength
$F^{\mu}{}_{\alpha\beta}$, that has a purely geometrical origin,
while the second one is related to the abelian gauge field
$\varphi_\alpha$ with the strength $\varphi_{\alpha\beta}$ and
appears due to quantum-mechanical nature of a theory (in the
classical limit $\hbar\to 0$ this object vanishes). As to terms in
(\ref{133}) with the factor $\hbar^{2}$, they describes rather the
operator orderings in the first terms in the r.~h.~s. of
(\ref{133}) than the features of dynamics.

\section{Hilbert Space of States}

Following the scheme introduced in \cite{ourwork1} we chose the set of
eigenvectors of coordinate operators $\{x^{M}\}$ as the  basis of the
Hilbert space for a point particle on $M$, namely

\begin{equation}
\hat{x}^{M}\left| x' \right\rangle=x'^{M}\left| x'
\right\rangle\,.
\label{h1}
\end{equation}
with the following normalization condition

\begin{equation}
\left\langle x'\right.\left| x'' \right\rangle= \Delta
(x'-x''):=\frac{1}{\sqrt{\eta(x)}}\delta(x'-x'')\,,
\label{h2}
\end{equation}
where $\eta(x)=\det(\eta_{MN}(x))$. The $\Delta$-function has the
following properties

\begin{equation}
f(x')\Delta (x'-x'')=f(x'')\Delta (x'-x'')\,,
\label{h3}
\end{equation}

\begin{equation}
f(x')\partial'_{M}\Delta (x'-x'')=
-\partial'_{M}f(x')\Delta (x'-x'')\,,
\label{h4}
\end{equation}

\begin{equation}
\partial'_{M}\Delta (x'-x'')=
\partial'_{M}\left( \frac{1}{\sqrt{\eta(x')}}\Delta (x'-x'') \right)=
-\Gamma_{M}(x')\Delta (x'-x'')-\partial''_{M}\Delta (x'-x'')\,.
\label{h5}
\end{equation}

In terms of coordinate decomposition $x^{M}=\{x^{\alpha} ,x^{\mu}\}$
the equation (\ref{h5}) takes the form

\begin{equation}
\left( \hat{D}'_{\alpha}+\hat{D}''_{\alpha}+
\hat{\Gamma}_{\alpha}(x')\right)\Delta (x'-x'')=
\partial'_{\mu}A^{\mu}_{\alpha}(x')\Delta (x'-x'')\,,
\label{h6}
\end{equation}
where

$$
\hat{D}_{\alpha}=\partial_{\alpha}-A^{\mu}_{\alpha}\partial_{\mu}\,,\quad
\hat{\Gamma}_{\alpha}=\Gamma_{\alpha}-A^{\mu}_{\alpha}\Gamma_{\mu}\,.
$$

To construct the coordinate representation of the operators describing the
quantum mechanics on $M$ one have to calculate the matrix elements of
coordinate and momentum operators $x^{M}$ and $p_{M}$.

In order to simplify this task we use previously defined operators

\begin{equation}
\overline{p}_{M}=\{\overline{p}_{\alpha},
\overline{p}_{\mu}\}\,,\quad
\overline{p}_{\alpha}:=p_{\alpha}-\varphi_{\alpha}\,,\quad
\overline{p}_{\mu}:=p_{\mu}\,,
\label{h7}
\end{equation}
that obey the following commutation relations

\begin{equation}
[x^{M}, x^{N}]=0\,,\quad
[x^{M}, \overline{p}_{N}]=i\hbar\delta^{M}_{N}\,,\quad
[\overline{p}_{M}, \overline{p}_{N}]=0\,.
\label{h8}
\end{equation}
Using the results of \cite{ourwork1} we can write

\begin{equation}
\begin{array}{rcl}
\left\langle x' \right|x^{M}\left| x'' \right\rangle&=& i\hbar
x'^{M}\left\langle x' \right.\left| x'' \right\rangle\,,\\[2mm]
\left\langle x' \right|\overline{p}_{M}\left| x'' \right\rangle&=&
-i\hbar\left(
\partial'_{M}+\dm\frac{1}{2}\Gamma_{M}(x')\right)\Delta
(x'-x'')\,.
\end{array}
\label{h8a}
\end{equation}

Returning to the operators $p_{M}$, we find that

\begin{equation}
\begin{array}{ccl}
\left\langle x' \right|p_{\alpha}\left| x'' \right\rangle &=&
-i\dm\hbar\left(
\partial'_{\alpha}+\dm\frac{1}{2}\Gamma_{\alpha}(x')\right)\Delta
(x'-x'')+ \varphi_{\alpha}(x')\Delta (x'-x'')\,,\\[2mm]
\left\langle x' \right|p_{\mu}\left| x'' \right\rangle &=&
-i\hbar\left( \partial'_{\mu}+\dm\frac{1}{2}\Gamma_{\mu}(x')
\right)\Delta (x'-x'')\,,\\[2mm] \left\langle x'
\right|x^{M}\left| x'' \right\rangle &=& x''^{M}\Delta (x'-x'')\,.
\end{array}
\label{h9}
\end{equation}

Using these expressions we can write the matrix element of the operator
$\pi_{\alpha}$ in the form

\begin{equation}
\left\langle x' \right|\pi_{\alpha}\left| x'' \right\rangle=
-i\hbar\left( \hat{D}_{\alpha}+
\frac{1}{2}(\hat{\Gamma}_{\alpha}-\partial_{\mu}A^{\mu}_{\alpha})\right)\Delta (x'-x'')+
\varphi_{\alpha}(x')\Delta (x'-x'')\,.
\label{h10}
\end{equation}

The coordinate representation of the operator $A$ can be constructed in terms
of wave functions $\psi(x')=\left\langle x' \right.\left| \psi \right\rangle $
representing the state $\left| \psi \right\rangle$ by the usual way:

\begin{equation}
(\hat{A}\psi)(x'):=
\int \left\langle x' \right|A\left| x'' \right\rangle
\left\langle x'' \right.\left| \psi \right\rangle d\,x''\,.
\label{h11}
\end{equation}
Therefore, following \cite{ourwork1} we find the representations
for the coordinate and momentum operators in the following form

\begin{equation}
\begin{array}{ccl}
\hat{x}^{M}\psi(x')&=&x'^{M}\psi(x')\,,\\[3mm]
\hat{p}_{\alpha}\psi(x')&=&-i\hbar\left( \partial'_{\alpha}+
\dm\frac{1}{2}\Gamma_{\alpha}(x')
\right)\psi(x')+\varphi_{\alpha}(x')\psi(x') \,,\\[3mm]
\hat{p}_{\mu}\psi(x')&=&-i\hbar\left(\partial'_{\mu}+\dm\frac{1}{2}\Gamma_{\mu}(x')
\right)\psi(x')\,.
\end{array}
\label{h12}
\end{equation}

The coordinate representation of the Hamiltonian can be obtained in terms of the
operators

\begin{equation}
\pi_{M}=p_{M}-\frac{i\hbar}{2}\Gamma_{M}\,,\quad
\pi^{\dag}_{M}=p_{M}+\frac{i\hbar}{2}\Gamma_{M}\,,
\label{h13}
\end{equation}
namely,

\begin{equation}
\hat{H}=\frac{1}{2}\hat{\pi}_{M}\hat{\eta}^{MN}\hat{\pi}^{\dag}_{N}\,.
\label{h14}
\end{equation}
Using the decomposition of the coordinates of $M$ we can write

\begin{equation}
\pi_{\alpha}=-i\hbar\left( \partial_{\alpha}+\Gamma_{\alpha}\right)+
\varphi_{\alpha}\,,\quad
\pi_{\mu}=-i\hbar\left( \partial_{\mu}+\Gamma_{\mu}  \right)\,,
\label{h15}
\end{equation}

\begin{equation}
\pi^{\dag}_{\alpha}=-i\hbar\partial_{\alpha}+\varphi_{\alpha}\,,\quad
\pi^{\dag}_{\mu}=-i\hbar\partial_{\mu}\,.
\label{h16}
\end{equation}

Hence, the formal part of the formulation of the quantum mechanics on the
Riemannian manifold with an intransitive group of isometries is solved.
Nevertheless, the other one, that contains the meaning and interpretation
of these results, requires the special discussion.

\section{Discussion}

We have examined in the present paper and in \cite{ourwork1},
\cite{ourwork2} the generalization of Schwinger's action principle for
the case of Riemannian manifolds with a group structure, determined by
isometric transformations.

The obtained results show fundamental unity between geometrical
properties of a manifold $M$ and the algebra of quantum-mechanical
operators, describing a point particle on $M$. The key role in the
formulation of quantum mechanics is played by geometrical aspects
of a symmetry, expressed in terms of Killing vectors. These
vectors also determine the integrals of motion on $M$ and their
algebraic properties.

It appears that a general quantization problem is conditionally reduced
into three cases. The first one we have considered in \cite{ourwork1}
is related to the class of manifolds with a simply transitive group of
isometries. Here the dimension of this group (which coincides with the
number of Killing vectors) is exactly equals to the dimension of the
manifold. The quantum mechanics receives its traditional form,
investigated in \cite{sugano}, \cite{dewitt}.

The second case contains the investigation of homogeneous manifolds
with non-simply transitive group of isometries, where the dimension of
a group exceeds the dimension of a manifold. There appears a
non-abelian gauge structure, that is similar to one in Kaluza-Klein
theories. This structure is related to the isotropy subgroup of $G$ and
plays the role of gauge group. This case has been analyzed in
\cite{ourwork2}. The obtained results are in accordance with
\cite{tsutsui}, where the method of study somewhat differs from ours.

The remaining third case, analyzed in the present paper, is
related to Riemannian manifolds with the intransitive group of
isometries $G$. In this case the dimension of the manifold $M$ is
bigger than the dimension of the group $G$. From the geometrical
point of view $M$ can be considered as a total space of the
principal fiber bundle with the base space $M/G$  and the
structure group $G$. In terms of this construction $M$ can be
covered by the set of submanifolds (the orbits of $G$-action) and
$G$ acts simply transitively in each of them. Quantum mechanics on
such a submanifold is developed in \cite{ourwork1}. The other
degrees of freedom, related to the quotient space $M/G$ have not
uniquely determined dynamics. In order to determine it
unequivocally we have to assume something new and external in
respect to Schwinger's scheme. There are two types of arbitrary
factors in the formulation of quantum mechanics on $M$. Some of
them have geometrical nature (the functions $b^{M}_{N}$), while
the others are quantum mechanical objects (the abelian gauge field
$\varphi_{\alpha}$ that vanishes when $\hbar\to 0$). It is
important to point out here, that the last structure can not be
obtained in the framework of canonical quantization procedure, it
is caused by geometrical properties of the manifold and algebraic
relation between geometrical objects.

The formal results of the present paper bring up the more
complicated question about its interpretation. We suppose that it
can be done in the following manner. There are some degrees of
freedom, describing a quantum system, which have partially
undetermined properties (as it can be treated in the traditional
meaning). The geometry of space is determined by energy-momentum
of the matter distribution. It is substantiated to search the
connection between geometrical structures corresponding to
undetermined dynamics and forbidden spatial zones for the system
with discrete energy spectrum (in the manner of ``layers'' between
the Bohr orbits in the hydrogen atom). Nevertheless, the question
about the interpretation of quantum theory on a general Riemannian
manifold remains an open one and requires more detailed discussion
and further development.

As the logical continuation of our series of papers we thought the
generalization of Schwinger's quantization approach on the case of a
supermanifold, that we are going to present in the forthcoming paper.


\begin{thebibliography}{99}

\bibitem{eisen}
L.~Eisenhart, {\it Continous Groups of Transformations},
Princeton, 1933

\bibitem{ourwork1}
N.~Chepilko, A.~Romanenko, Submitted to Europ. Phys. Journ.

\bibitem{ourwork2}
N.~Chepilko, A.~Romanenko, Submitted to Europ. Phys. Journ.

\bibitem{sugano}
R. Sugano, {\it Prog.Theor.Phys.} {\bf 46}, 297 (1971);\newline
T. Kimura, {\it ibid.} {\bf 46}, 126 (1971);\newline
R. Sugano and T. Kimura, {\it ibid.} {\bf 47}, 1004 (1972);\newline
T. Ohtani and R. Sugano, {\it ibid.} {\bf 47}, 1704 (1972);
{\it ibid.} {\bf 50}.
1705 (1973)

\bibitem{dewitt}
B.~S.~DeWitt, {\it Rev. Mod. Phys.} {\bf 29}, 377 (1957).

\bibitem{tsutsui}
 D. McMullan and I. Tsutsui, {\it Ann. Phys.} {\bf 237}, 269 (1995);
\newline
 Y. Ohnuki and S. Kitakado {\it J. Math. Phys.} {\bf 34}, 2827 (1993).
\end{thebibliography}
\end{document}